\documentclass[pra,showpacs,twocolumn]{revtex4}
\usepackage{bm,graphicx,amsmath}
\begin{document}
\title{Dynamical quantum phase transition of a two-component Bose-Einstein condensate in an optical lattice}
\author{Anssi Collin, Jani-Petri Martikainen, and Jonas Larson}
\email{jolarson@kth.se} \affiliation{NORDITA, 106 91 Stockholm,
Sweden}

\date{\today}

\begin{abstract}
We study dynamics of a two-component Bose-Einstein condensate where
the two components are coupled via an optical lattice. In
particular, we focus on the dynamics as one drives the system
through a critical point of a first order phase transition
characterized by a jump in the internal populations. Solving the
time-dependent Gross-Pitaevskii equation, we analyze; breakdown of
adiabaticity, impact of non-linear atom-atom scattering, and the
role of a harmonic trapping potential. Our findings demonstrate that
the phase transition is resilient to both contact interaction
between atoms and external trapping confinement.
\end{abstract}
\maketitle

\section{Introduction}
Since the pioneering experiments on Bose-Einsten condensates (BEC)
\cite{bec}, the field of ultracold atomic gases has seen a
tremendous development \cite{bec2}. Nowadays, preparation and
manipulation of BECs is a standard procedure providing a
versatile testbed for the study of various quantum effects. Spinor
condensates, where internal atomic Zeeman levels play an important
role for the dynamics, have been experimentally studied by
numerous groups \cite{spinor}. More recent experiments on spinor
condensates include; coherent transport \cite{spintrans},
spin-mixing \cite{spinmix}, inherent spin tunneling \cite{spintun},
and symmetry breaking \cite{spinbreak}. Placing the condensate in an
optical lattice formed by counter propagating laser beams greatly
affects its characteristics and exciting phenomena, {\it e.g.}
Bloch oscillations \cite{bloch}, gap and colliding solitons \cite{gapsol,solcol}, self-trapping \cite{selftrap},
superfluid instability \cite{super}, vortices \cite{vortex},
Anderson localization \cite{anderson}, and for more strongly
correlated systems the superfluid-Mott insulator phase
transition (PT) \cite{mott}, arise.

Most theoretical works studying PTs consider systems at strict
thermal equilibrium. However, a more appropriate picture capturing
the underlying physics of real experiments is often encountered by a
dynamical approach. Indeed, the great experimental progress seen in
the recent past on correlated ultracold atomic systems
\cite{maciek}, calls for a deeper theoretical understanding of
non-equilibrium PTs. Much of the theoretical works are devoted to
dynamics across a critical point in various spin models
\cite{spintr0,spintr1,spintr2,spintr3,spintr4}. Also the dynamics of
phase transitions in lattice many-body systems such as the
Bose-Hubbard model have been considered \cite{kzbh1,kzbh2}. To
derive an estimate of the adiabatic breakdown, the Landau-Zener
formula, giving a measure of transition probabilities across an
avoided level crossing \cite{lz}, has been utilized in most of the
above works \cite{spintr0,spintr1,spintr2,spintr3,kzbh1}. Moreover,
in Refs.~\cite{spintr0,spintr1,spintr2,kzbh1,kzbh2} the Kibble-Zurek
mechanism of quench induced transitions was employed.

In this paper we consider a spinor condensate in an optical lattice.
Apart from providing an effective periodic potential, in this model
the lattice also induces a coupling between the internal atomic
states. The corresponding many-body model, without atom-atom
interactions, was first introduced in Ref.~\cite{jonasjani}, and the
occurrence of a first order PT between different internal states was
demonstrated. The system was further investigated in
\cite{jonasjani2} by taking interaction between the atoms into
account. Both of these works consider thermodynamical equilibrium.
Since the single particle ground state of the present model is not
gapped, even small time-dependent perturbations of the Hamiltonian
might well cause non-adiabatic excitations. Driven through the
critical point, such excitations might wash out the signatures of
the PT. One goal of the present paper is to study the importance of
these non-adiabatic contributions. Moreover, the model automatically
takes into account the effects deriving from atom-atom interactions,
and the state changing collisions they bring about. This aspect was
left as an open question in \cite{jonasjani}. In addition, the
sensitivity to external harmonic trapping is also considered.

The outline of the paper is as follows. In the next section, we
present the one dimensional single particle Hamiltonian. After first discussing
some general properties in Subsec.~\ref{ssec2a}, we
numerically diagonalize the Hamiltonian and give the spectrum in
Subsec.~\ref{ssec2b}. Using the spectrum, we demonstrate the
equilibrium PT in Subsec.~\ref{ssec2c}. The following
Sec.~\ref{sec3} is devoted to our main results; the dynamics when
the system is driven through the critical point. The numerical
method for solving the Gross-Pitaevskii equations is outlined, and
first the thermal equilibrium PT is studied by means of the
partial-state fidelity susceptibility. In Subsec.~\ref{ssec3a} we
calculate the order parameter for different dimensions, whereby the
following Subsection considers the situation of external harmonic
trapping. Breakdown of adiabaticity is analyzed in
Subsec.~\ref{ssec3c}. Finally we conclude in Sec.~\ref{sec4}.

\section{One dimensional ideal gas model system}\label{sec2}
The appearance and nature of the PT are easily extracted from the
structure of the spectrum of the one dimensional single particle
Hamiltonian. The effects of particle interactions, higher
dimensional systems, and  explicit time-dependence of system parameters
will be considered in Sec.~\ref{sec3}.

\subsection{Ideal gas Hamiltonian}\label{ssec2a}
\begin{figure}[ht]
\begin{center}
\includegraphics[width=8cm]{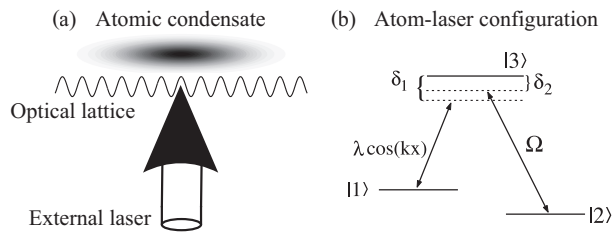}
\caption{Schematic configuration of system setup (a) and atom-laser
configuration (b). An optical lattice drives the
$|1\rangle\leftrightarrow|3\rangle$ atomic transition and the
``external" laser the $|2\rangle\leftrightarrow|3\rangle$ transition. }
\label{fig1}
\end{center}
\end{figure}

The properties of an ideal gas are captured in the corresponding
single particle system. We consider a three-level
$\Lambda$ atom dipole coupled to two light fields as illustrated in
Fig.~\ref{fig1}. The $1\leftrightarrow3$ transition is driven via a
periodic standing wave field, while an ``external" spatially
independent laser gives rise to the $2\leftrightarrow3$ transition.
This assumes the perpendicular laser to have a mode-waist larger
than the extent of the condensate. The effective atom-field
couplings are $\lambda$ and $\Omega$, the wave number $k$, and the
atom-field detunings $\delta_1$ and $\delta_2$ respectively. In the
case of large detunings, the excited atomic state $|3\rangle$ may be
adiabatically eliminated yielding an effective $2\times2$
Hamiltonian \cite{adel}
\begin{equation}\label{ham1}
\hat{H}_1=\frac{\hat{\tilde{p}}^2}{2m}+\frac{\hbar\tilde{\Delta}}{2}\hat{\sigma}_z-\hbar\tilde{U}_1\cos(2k\hat{x})\hat{\sigma}_{11}+\hbar\tilde{U}\cos(k\hat{x})\hat{\sigma}_x,
\end{equation}
where
$\tilde{\Delta}=|\delta_1-\delta_2|-\Omega^2/\delta_2-\lambda^2/2\delta_1$
is an effective detuning taking into account for the constant Stark
shifts of states 1 and 2, $\tilde{U}_1=\lambda^2/2\delta_1$,
$\tilde{U}=\lambda\Omega(1/2\delta_1+1/2\delta_2)$, and the
$\sigma$-operators are the Pauli matrices;
$\hat{\sigma}_z=|2\rangle\langle2|-|1\rangle\langle1|$,
$\hat{\sigma}_x=|1\rangle\langle2|+|2\rangle\langle1|$, and
$\hat{\sigma}_{11}=|1\rangle\langle1|$. Respectively,
$\hat{\tilde{p}}$, $\hat{\tilde{x}}$, and $m$ are atomic momentum,
position and mass. Note that the $\tilde{U}_1$ term originates from
the Stark shift of the $|1\rangle$ state. As it affects only the
$|1\rangle$ state and not the $|2\rangle$ state, it shifts the
detuning resonance
$\tilde{\Delta}_c=0\rightarrow\tilde{\Delta}_c\neq0$. Apart from
altering the detuning resonance, it also renders a light
induced potential for the state $|1\rangle$. The $\tilde{U}$ term,
on the other hand, provides an effective potential for both internal
states. In the limiting cases, when one of the parameters
$\tilde{\Delta}$, $\tilde{U}_1$ and $\tilde{U}$ is dominating the
other two, it is legitimate to assign a particular potential to the
internal atomic states. In general, however, the full coupled system
has to be considered as an entity \cite{jonasjani2}.

For later convenience, scaled quantities are introduced by the
characteristic energy and length scales $E_r=\frac{\hbar^2k^2}{2m}$
and $k^{-1}$, such that
\begin{equation}
\hat{x}=k\hat{\tilde{x}},\hspace{0.6cm}\Delta=\frac{\hbar\tilde{\Delta}}{E_r},\hspace{0.6cm}U_1=\frac{\hbar\tilde{U}_1}{E_r},\hspace{0.6cm}U=\frac{\hbar\tilde{U}}{E_r}.
\end{equation}
In the matrix-form, the scaled Hamiltonian reads
\begin{equation}\label{ham2}
\hat{H}_1=-\frac{\partial^2}{\partial x^2}+\left[\begin{array}{cc}
\displaystyle{\frac{\Delta}{2}} & U\cos(\hat{x}) \\
U\cos(\hat{x}) & -\displaystyle{\frac{\Delta}{2}}-U_1\cos(2\hat{x})\end{array}\right],
\end{equation}
where we have $|1\rangle=\left[\begin{array}{c}0\\
1\end{array}\right]$and $|2\rangle=\left[\begin{array}{c}1\\
0\end{array}\right]$. From the fact that
$\cos(\hat{x})|q\rangle=(|q+1\rangle+|q-1\rangle)/2$, where
$\hat{p}|q\rangle=q|q\rangle$, one finds that a swapping of the
internal states 1 and 2 is accompanied by a shift of momentum by
one unit in either positive or negative direction. Moreover, a given
momentum eigenstate $|q\rangle$ is only coupled to other momentum
states $|q+\eta\rangle$, where $\eta$ is any integer. In particular,
the states $\{|q\rangle|i\rangle\}$ ($i=1,2$) can be divided into
two sets
\begin{equation}\label{basis}
\begin{array}{l}
|\varphi_\eta(q)\rangle=\left\{\begin{array}{l}
|q+\eta\rangle|2\rangle\hspace{1cm}\eta\,\,\mathrm{even}\\
|q+\eta\rangle|1\rangle\hspace{1cm}\eta\,\,\mathrm{odd}\end{array}\right.\\ \\
|\phi_\eta(q)\rangle=\left\{\begin{array}{l}
|q+\eta\rangle|1\rangle\hspace{1cm}\eta\,\,\mathrm{even}\\
|q+\eta\rangle|2\rangle\hspace{1cm}\eta\,\,\mathrm{odd},\end{array}\right.
\end{array}
\end{equation}
which are not coupled by the Hamiltonian (\ref{ham2});
$\langle\varphi_{\eta'}(q')|\hat{H}_1|\phi_\eta(q)\rangle=0$. Thus,
the Hamiltonian may be written as
$\hat{H}_1=\hat{H}_\varphi\otimes\hat{H}_\phi$, where the two
sub-Hamiltonians operate on states spanned by
$\{|\varphi_\eta(q)\rangle\}$ or $\{|\phi_\eta(q)\rangle\}$
respectively. Here, the quasi momentum $q$ resides in the first
Brillouin zone.

As a periodic operator, $\hat{H}_1$ commutes with $\hat{T}=e^{\pm
il\hat{p}}$, where $l=2\pi$ is the scaled period. It is readily
shown \cite{jonaseff} that the operator
\begin{equation}
\hat{I}=\hat{\sigma}_z\mathrm{e}^{\pm i\frac{l}{2}\hat{p}},
\end{equation}
defines another symmetry, describing a half-period boost of the
position and an atomic inversion. An outcome of this $l/2$ symmetry
is that the Brillouin zone extends beyond $(-\pi/l,\pi/l]$ as
imposed by the $l$-periodicity of the system
\cite{jonasjani,jonaseff}. Thereby, the first Brillouin zone is
instead defined for quasi momentum $q\in(-1,1]$.

\subsection{Ideal gas energy spectrum}\label{ssec2b}
As was already pointed out, the character of the PT is directly seen
from the energy spectrum. Due to the block diagonal form of the
Hamiltonian, $\hat{H}_1=\hat{H}_\varphi\otimes\hat{H}_\phi$, the
dispersions of $\hat{H}_\varphi$ and $\hat{H}_\phi$ can be
calculated separately. In the figures, we will differentiate between
dispersions of $\hat{H}_\varphi$ and $\hat{H}_\phi$ by using dashed
or solid curves, respectively.

The internal two-level structure of the system brings about
anomalous dispersions that possess several minima
\cite{jonasjani2,konstantin}. The lowest energy of each band
$E_\nu(q)$, where $\nu=1,\,2,\,3,\,...$ is the band index and
$q\in(-1,1]$ the quasi momentum, is twofold degenerate. However, the
degeneracy is due to the fact that the individual energy bands
$E_\nu^{(\varphi)}(q)=E_\nu^{(\phi)}(q')$ for some $q$ and $q'$. The
only exception is at exact resonance $\Delta=\Delta_c$, where the
bands $E_\nu^{(\varphi)}(q)$ and $E_\nu^{(\phi)}(q)$ become
identical.

\begin{figure*}[ht]
\begin{center}
\includegraphics[width=14cm]{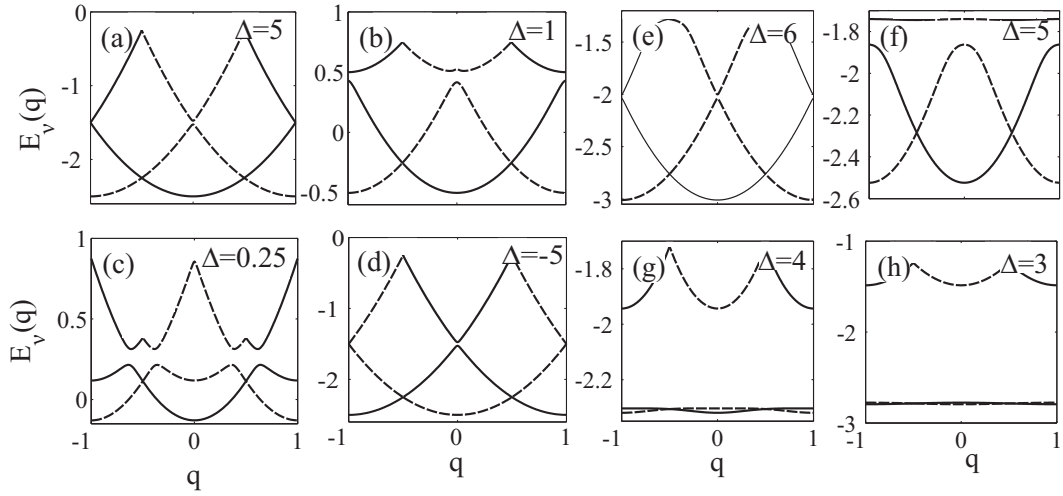}
\caption{Examples of the three lowest energy bands of Hamiltonian
(\ref{ham2}). In (a)-(d) we have $U>U_1$ ($U=0.1$ and $U_1=0.05$),
while in (e)-(h) $U<U_1$ ($U=0.2$ and $U_1=8$). The multiple number
of local minima of the lowest band is clear. Dashed lines are the
energy dispersions $E_\nu^{(\varphi)}(q)$ of the sub-Hamiltonian
$\hat{H}_\varphi$, and solid lines are $E_\nu^{(\phi)}(q)$
corresponding to $\hat{H}_\phi$. } \label{fig2}
\end{center}
\end{figure*}

The Hamiltonian (\ref{ham2}) is readily diagonalized by truncating
the number of basis states (\ref{basis}). If $U_1\gg U$, atoms in
the $|1\rangle$ state will feel a strong periodic potential, while
$|2\rangle$ atoms experience only a weak potential. Thus, the
spectrum is in this case typically made up of a fairly flat band
mixed together with ``parabolic" energy bands. When $U$ dominates
$U_1$, the situation is different and mixing of flat and
``parabolic" bands does not exist. Examples of the first three
energy dispersions are displayed in Fig.~\ref{fig2}. In (a)-(d)
$U>U_1$, while in (e)-(h) $U_1\gg U$. In the last three plots,
almost flat dispersions are found due to the large $U_1$, and
especially in (f) there are ``parabolic" bands with lower energy
than the flat one.

\subsection{First order phase transition}\label{ssec2c}
At zero temperature, the condensate atoms will occupy the ground
state. Even though the ground state is degenerate, mainly one of the
degenerate states will be populated. In an experiment, the atoms
will typically be initialized in either $|1\rangle$ or $|2\rangle$
and then cooled down to its ground state. Hence, the obtained
initial ground state will predominantly populate either of the two
energy bands $E_1^{(\varphi)}(q)$ or $E_1^{(\phi)}(q)$. An equal
population balance between the degenerate ground states is very
unlikely, and indeed, in our numerical simulations of interacting
systems utilizing the imaginary time propagation method we never
encounter such balanced situations.

Comparing Fig.~\ref{fig2} (a) and (d), it is seen that the
dispersions $E_\nu^{(\varphi)}(q)$ and $E_\nu^{(\phi)}(q)$ have been
roughly swapped when $\Delta\leftrightarrow-\Delta$. Thus, the
minimum of $E_1^{(\varphi)}(q)$ or $E_1^{(\phi)}(q)$ are changed
between the center and edges of the Brillouin zone as $\Delta$ is
tuned through the resonance $\Delta=\Delta_c$. In the example of
Fig.~\ref{fig2} (a)-(d), this resonance condition is
$\Delta_c\approx0$ since $U_1\ll1$. When $U_1$ is large, however,
the $\Delta_c$ is shifted away from zero, but nonetheless, between
$\Delta<\Delta_c$ and $\Delta>\Delta_c$ the minima is either at the
Brillouin center or at its edges. As the quasi momentum
corresponding to the energy minima of either $E_1^{(\varphi)}(q)$ or
$E_1^{(\phi)}(q)$ changes abruptly when $\Delta$ is tuned across the
resonance, one encounters a first order PT.

\begin{figure}[h]
\centerline{\includegraphics[width=7cm]{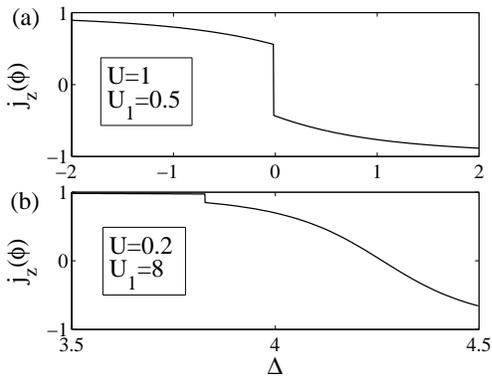}} \caption{The
inversion (\ref{inv}) as function of detuning $\Delta$. In the upper
plot, $U_1\ll1$ and $\Delta_c\approx0$, while in the lower plot
$U_1=8$ giving $\Delta_c\approx3.8$.  } \label{fig3}
\end{figure}

As discussed in the previous subsection, the internal ($|1\rangle$
and $|2\rangle$) and external (momentum) degrees of freedom cannot
be changed independently via the Hamiltonian (\ref{ham2}).
Absorption or emittence of a lattice photon always induces a
momentum ``kick''. Consequently, shifting the momentum of the ground
state between the center and the edges of the Brillouin zone is only
possible by swapping the internal states $|1\rangle$ and
$|2\rangle$. Thus, the collective inversion for atoms belonging to
either of the $E_1^{(\varphi)}(q)$ or $E_1^{(\phi)}(q)$ bands
\begin{equation}\label{inv}
j_z^{(\gamma)}=\frac{1}{N_\gamma}\sum_{i=1}^{N_\gamma}\langle\hat{\sigma}_z^{(i)}\rangle,\hspace{1.2cm}\gamma=\varphi,\,\phi,
\end{equation}
where $\hat{\sigma}_z^{(i)}$ is the $i$th atom's $z$ Pauli matrix,
should be discontinuous at $\Delta_c$. Here, $N_\alpha$ is the
number of condensate atoms in band $E_1^{(\gamma)}(q)$, and we note
that since we consider only the ground state,
$\langle\hat{\sigma}_z^{(i)}\rangle=\langle\hat{\sigma}_z^{(j)}\rangle$
for any $i$ and $j$. The full inversion
$j_z=j_z^{(\varphi)}+j_z^{(\phi)}$, however, depends on the number
of atoms residing in each band.

Figure~\ref{fig3} displays two examples of $j_z$ when all atoms
belong to the $E_1^{(\phi)}(q)$ band. In (a), the inversion is shown
for the parameters corresponding to Figs.~\ref{fig2} (a)-(d), and
the gap of the inversion is found to be well resolved. In (b),
however, corresponding to Figs.~\ref{fig2} (e)-(h), reveals a small
gap in the inversion. In this case, the PT occurs for a detuning
$\Delta_c\approx3.8$ which is somewhere between Fig.~\ref{fig2} (g)
and (h) where the dispersions are fairly flat. We note that apart
from shifting the critical resonance $\Delta_c$, large $U_1$'s also
results in asymmetric $j_z^{(\phi)}$'s with respect to $\Delta_c$.

\section{Dynamical phase transition of a weakly interacting
Bose-Einstein condensate}\label{sec3} Section \ref{sec2}
demonstrated that a first order PT is encountered by driving the
system of non-interacting atoms through a critical detuning
$\Delta_c$. Adding interaction between the atoms will introduce
coupling between the two sets of basis states (\ref{basis}). This
effect is, however, relatively small~\cite{jonasjani2}. On the other
hand, very strong atom-atom scattering amplitudes can have other
important consequences, for example it may drastically change the
single particle dispersion curves \cite{pethick}. Nonetheless, it
was argued in Refs.~\cite{jonasjani,jonasjani2}, that the PT should
be stable even against strong interaction between the particles. In
this section we will show, on a mean-field basis, that this is
indeed true, and that the PT is also insensitive to an external
harmonic trapping potential. Moreover, the dynamics of the
condensate driven through the critical point will be thoroughly
addressed.

\subsection{Sweep through the critical point}\label{ssec3a}

\begin{figure}[h]
\centerline{\includegraphics[width=7cm]{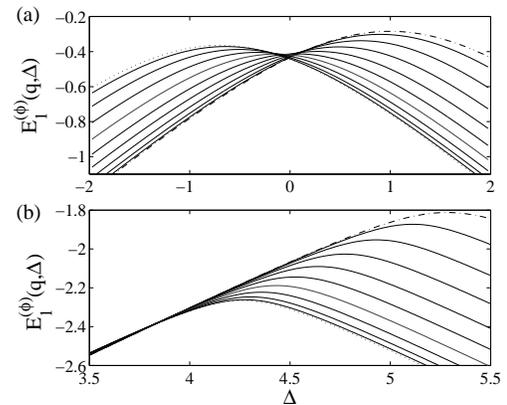}} \caption{Examples
of the lowest band energies $E_1^{(\phi)}(q_n,\Delta)$ for
$q_n=-n/10$, $n=0,\,1,\,...\,10$. The parameters are the same as in
Fig.~\ref{fig3}; $U=1$ and $U_1=0.5$ in (a), and $U=0.2$ and
$U_1=8$ in (b). The dotted line and the dot-dashed line are the
energies $E_1^{(\phi)}(-1,\Delta)$ and $E_1^{(\phi)}(0,\Delta)$
respectively, and the crossing between these two,
$E_1^{(\phi)}(0,\Delta)=E_1^{(\phi)}(-1,\Delta)$, gives the critical
detuning $\Delta_c$. } \label{fig4}
\end{figure}

The spectra $E_1^{(\varphi)}(q=0,\Delta)$ and
$E_1^{(\varphi)}(q=\pm1,\Delta)$, and the spectra
$E_1^{(\phi)}(q=0,\Delta)$ and $E_1^{(\phi)}(q=\pm1,\Delta)$ become
degenerate and cross at $\Delta=\Delta_c$. Such crossing of energies
is characteristic for a PT \cite{sachdev}. The crossing of energies
in the present model is envisaged in Fig.~\ref{fig4} displaying the
energies $E_1^{(\phi)}(q,\Delta)$ as function of $\Delta$ for
$q=-n/N$, where $N=10$ and $n=0,\,1,\,...\,N$. Letting
$N\rightarrow\infty$, the spacings between the lines in
Fig.~\ref{fig3} vanish. The two figures use the same parameters as
in Fig.~\ref{fig3}, and we note that when $U_1$ is large, $\Delta_c$
is shifted away from zero as was already seen in Fig.~\ref{fig3}.
Further, in (a) corresponding to Fig.~\ref{fig3} (a) where the
discontinuity of the order parameter is more pronounced, the
energies are more sparsely spaced compared to (b). Indeed, in (b) it
is not even clear that the energies $E_1^{(\phi)}(q=0,\Delta)$ and
$E_1^{(\phi)}(q=\pm1,\Delta)$ actually cross. However, a closer look
around the critical point reveals the crossing.

The purpose of this section is to study the effects on the PT
arising from interaction between the particles and from a confining
harmonic trap. Furthermore, the analysis will not be restricted to
the one dimensional case, but we also considers the two dimensional
counterparts. One interesting aspect is the dynamics in itself as
the system is driven through the critical point. Another central
aspect that will be investigated is breakdown of adiabaticity. Due
to the continuous spectrum and the crossing of the energy levels,
non-adiabatic excitations of the condensate are expected and
especially around $\Delta=\Delta_c$ where the density of states
drastically increases.

In the higher dimensional cases we assume, for simplicity, that all
lattice fields couple with the same strength to the atom and that
they share the same periodicity. Such a scheme requires multiple
excited Zeeman levels as well as mutually orthogonal polarizations
of the lattice beams, but after the adiabatic elimination on all of
them, one is left with an effective two-level Hamiltonian
\begin{equation}\label{ham3}
\hat{H}=\!-\hat{\nabla}^2+\!\left[\begin{array}{cc}
\displaystyle{\frac{\Delta}{2}} & U\displaystyle{\sum_{\alpha}\cos(x_\alpha)} \\
U\displaystyle{\sum_{\alpha}}\cos(x_\alpha) & -\displaystyle{\frac{\Delta}{2}}\!-U_1\!\sum_{\alpha}\cos^2(x_\alpha)\end{array}\!\!\right]\!.
\end{equation}
Here $\alpha$ indicates a component of the vector along the direction $\alpha$.

Moreover, we will take into account for collisions between the
particles in terms of a non-linear term in the Hamiltonian, as well
as consider a harmonic trapping potential. We thus consider the
Gross-Pitaevskii equation
\begin{equation}\label{gp1}
\begin{array}{lll}
i\displaystyle{\frac{\partial}{\partial t}\Psi(\mathbf{x},t)}
&  = & \hat{H}_{GP}\Psi(\mathbf{x},t)\\ \\
& = &
\displaystyle{\left[\hat{H}\!+\!\frac{\omega^2}{2}\!\sum_\alpha x_\alpha^2+\Psi^\dagger(\mathbf{x})\!\cdot\!\mathbf{g}\!\cdot\!\Psi(\mathbf{x})\right]\!\!\Psi(\mathbf{x},t)},
\end{array}
\end{equation}
where $t$ is the scaled time, $\omega$ the scaled parameter
determining the trap frequency, and $\bf{g}$ the scattering
amplitudes. In this coupled two-level problem, the wave functions are
spinors
\begin{equation}
\Psi(\mathbf{x})=\left[\begin{array}{c}\psi_2(\mathbf{x})\\ \psi_1(\mathbf{x})\end{array}\right]
\end{equation}
and
\begin{equation}
\mathbf{g}=\left[\begin{array}{cc} g_{11} & g_{12} \\
g_{12} & g_{22}\end{array}\right].
\end{equation}
Normalizing the wave function as $\int
d\mathbf{x}|\Psi(\mathbf{x})|^2=1$, the scattering amplitudes become
$g_{ij}=4\pi\hbar Na_{ij}/mVE_r$, where $N$ is the number of
condensate atoms, $V$ the effective volume, and $a_{ij}$ the state
dependent $s$-wave scattering amplitude.

We will solve Eq.~(\ref{gp1}) numerically, starting
in the ground state for large detuning and then change $\Delta$ in
time such that it passes the critical point. In particular we chose a
linear sweep of the detuning,
\begin{equation}\label{timedet}
\Delta(t)=\displaystyle{\Delta_0t},
\end{equation}
where $\Delta_0$ sets the sweep velocity. We integrate (\ref{gp1})
from large negative to large positive times $t\in[-T,T]$ such that
the critical point is surely traversed.

The GP equation (\ref{gp1}) is solved using the split operator wave
packet method \cite{split}. The initial state is taken as the ground
state for the given system parameters. In particular, the ground
state is obtained via the imaginary time evolution
\begin{equation}\label{gs}
\Psi_{GS}(\mathbf{x})=\lim_{\tau\rightarrow\infty}\frac{\Psi_\mathbf{0}(\mathbf{x},\tau)}{\sqrt{\int\,d\mathbf{x}|\Psi_\mathbf{0}(\mathbf{x},\tau)|^2}},
\end{equation}
where
\begin{equation}\label{ansatz}
\Psi_\mathbf{0}(\mathbf{x},\tau)=e^{-\tau\hat{H}_{GP}}\Psi_\mathbf{0}(\mathbf{x},0)\left[\begin{array}{c}1 \\ 0\end{array}\right]
\end{equation}
and $\Psi_\mathbf{0}(\mathbf{x},0)$ is the initial wave function
ansatz which will be taken as a normalized real Gaussian. The
subscript $\mathbf{0}$ indicates the initial Hamiltonian parameters
$U$, $U_1$, $\mathbf{g}$, $\Delta_0$, and $T$, as well as the center
of the Gaussian $\mathbf{x}_0=0$ and its width $\Delta_\mathbf{x}$
(which is taken to be much larger than the period $l=2\pi$ of the
Hamiltonian). This numerical procedure gives a good approximation
for the ground state. Another important point is that since the
ground state (for a non-interacting system) is degenerate, the
imaginary time evolution renders different states depending on the
ansatz (\ref{ansatz}). In particular, since for the state of our
assumption all population resides in the $|1\rangle$ state and the
momentum is centered around the origin, the ground state obtained
from (\ref{gs}) will mainly populate states spanned by
$\{|\varphi_\eta(q)\rangle\}$. This holds even for relatively large
$\bf{g}$. Again, an imbalanced initial state is believed to be in
good agreement with true experimental realizations; roughly
speaking, the imaginary time evolution represents the cooling of the
condensate.

\begin{figure}[h]
\centerline{\includegraphics[width=7cm]{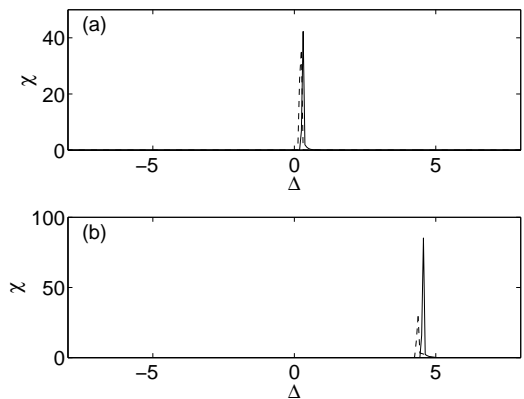}}
\caption{The partial-state fidelity susceptibility (\ref{sus}) for
the one dimensional model. In (a) $U=1$ and $U_1=0.5$,
while in (b) $U=0.2$ and $U_1=8$. In both plots $\omega=0$,
 and $g_{11}=g_{22}=g_{12}=0$ (solid lines),
$g_{11}=g_{22}=10g_{12}=10$ (dashed lines). } \label{fig5}
\end{figure}

Before studying the actual dynamic problem, we consider the PT
characterized by the ground states numerically obtained via
imaginary time evolution (\ref{gs}). By doing so, we get an idea of
how sensitive the PT is to the non-linear atom-atom interaction
term. In particular, we evaluate the partial-state fidelity
susceptibility, which has been demonstrated to give a clear
indication of various PTs \cite{fidde0}. The advantage of utilizing
a fidelity measure, originating from the quantum information
concepts, to analyzing PTs is that it is state independent and no
prior knowledge about order parameters is required.
 The fidelity susceptibility has been shown to contain
sufficient information to reveal the universality class of the PTs
\cite{fidde1}, even for topological PTs \cite{topfid},
Kosterlitz-Thouless transitions \cite{fidde1}, and PTs characterized
by an underlying continuous level crossing \cite{fidde2}. The idea
is simple, letting the ground state be parameterized by some
quantity $h$ (in our case $h=\Delta$),
$\rho(h)=|\Psi(h)\rangle\langle\Psi(h)|$ we introduce the reduced
density operator for the internal states
$\rho_A(h)=\mathrm{Tr}_B\left[\rho(h)\right]$, where the trace is
over external degrees of freedom. For the present model, it is
straightforward to derive
\begin{equation}
\rho_A(h)=\left[\begin{array}{cc} N_{11}(h) & N_{12}(h) \\
N_{21}(h) & N_{22}(h)\end{array}\right],
\end{equation}
where
$N_{ij}=\int\,d\mathbf{x}\psi_i^*(\mathbf{x},h)\psi_j(\mathbf{x},h)$
and $\psi_i(\mathbf{x},h)$ is the constituent ground state wave
function for the parameter $h$. The partial-state fidelity is
defined as
\begin{equation}\label{partfid}
F_A(h,\delta h)=\mathrm{Tr}\left[\sqrt{\sqrt{\rho_A(h)}\rho_A(h+\delta h)\sqrt{\rho_A(h)}}\right].
\end{equation}
In terms of $F_A(h,\delta h)$, the partial-state fidelity susceptibility reads
\begin{equation}\label{sus}
\chi(h)=\lim_{\delta h\rightarrow0}\frac{-2\ln F_A(h,\delta h)}{\delta h^2}.
\end{equation}
For the situation at hand, there is a critical $h_c(=\Delta_c)$ for
which the PT occurs. The fidelity measures the overlap between the
groundstates for different parameters $h$ and $h+\delta h$, and
whenever $h_c$ lies between these two values the fidelity will
drastically drop and the susceptibility increase. This is displayed
in Fig.~\ref{fig5}. The upper plot (a) represent the case with
$U>U_1$, while (b) shows examples of the reverse situation. For
solid lines, the atom-atom interaction is zero and for dashed lines
it is non-zero. The sharp peaks of the susceptibility at $h_c$
coincide well with the critical parameters found in Fig.~\ref{fig3}
using the full numerical diagonalization of the single particle
Hamiltonian. The slight discrepancy is believed to derive from the
fact that the imaginary time propagation is stopped after some
finite time $\tau$. Interestingly, the shape and location of the
peaks of $\chi(h)$ seems to be insensitive for atom-atom
interactions, proving that the PT is not restricted to an ideal gas.

\begin{figure}[h]
\centerline{\includegraphics[width=7cm]{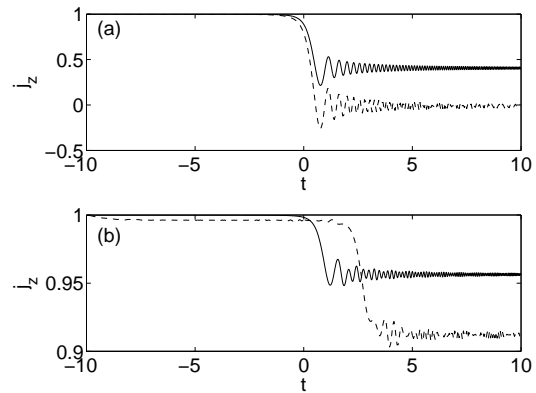}} \caption{Examples
of the collective inversion $j_z$ as function of time; one dimension
(solid lines) and two dimensions (dashed lines). In (a) $U=1$ and
$U_1=0.5$, while in (b) $U=0.2$ and $U_1=8$. In both plots
$\Delta_0=10$, $\omega=0$, $g_{11}=g_{22}=g_{12}=0$. } \label{fig6}
\end{figure}

Let us now turn to dynamics. Once the ground state is initialized,
the wave packet is let to evolve using the real time propagator.
Since the Hamiltonian is explicitly time-dependent, the evolution is
non-trivial. Figure \ref{fig6} shows examples of the collective
inversion $j_z=j_z^{(\phi)}+j_z^{(\varphi)}$, corresponding to the
parameters of Fig.~\ref{fig3}. Solid lines give the one dimensional
case and dashed lines the two dimensional case. The figure makes
clear that the PT is resilient to the dimensionality of the system.
In fact, the change in the inversion is even more pronounced in the
two dimensional situation. The shift of the critical point between
the two dimensions seen in (b) can be understood from the
Hamiltonian (\ref{ham3}). Since in two dimensions the sum on the
diagonal contains two terms and both terms shift the resonance
condition, the shifting is larger for higher dimensions.

\subsection{Effect of a trapping potential}\label{ssec3b}
The PT demonstrated in the previous section is an outcome of the
periodicity of our system and in particular the anomalous energy
dispersions. The band structure of the spectrum will be lost if the
condensate is placed in an additional external trapping potential.
For weak confinement, however, the spectrum still shows similarities
to a band-gap spectrum. More precisely, if the characteristic length
scale of the harmonic trap is much larger than the lattice
periodicity $l=2\pi$, signatures of the PT are supposedly still
present. On the other hand, very tight trapping may wash out any signatures of
the PT.

\begin{figure}[h]
\centerline{\includegraphics[width=7cm]{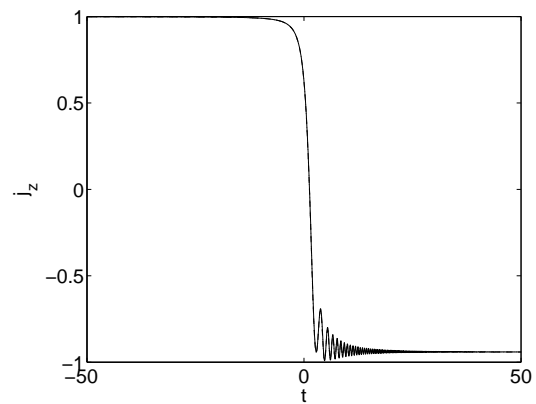}}
\caption{The collective inversion $j_z$ as function of time for the
one dimensional case. Solid line ($\omega=0.01$), dashed line
($\omega=0.02$) and dotted line ($\omega=0.05$). The rest of the
dimensionless parameters are, $U=1$, $U_1=0.5$, $\Delta_0=5$, and $g_{11}=g_{22}=g_{12}=0$.}
\label{fig7}
\end{figure}

In Fig.~\ref{fig7} we display examples of the collective inversion
for three different trapping strengths. For these parameters the
three lines are almost identical, and still for the largest $\omega$
the ground state wave function only extends over roughly five
lattice periods. By increasing $\omega$ further one cannot talk
about a periodic system, and it is found that the momentum wave
packet does not undergo a sudden momentum shift when driven through
the critical point. The atom-scattering $\mathbf{g}=0$ in these
examples.

\subsection{Breakdown of adiabaticity}\label{ssec3c}
As already pointed out, the ground state is not gapped in the
current system, which should make it sensitive to non-adiabatic
effects. Adiabaticity is still most likely to break down at the
critical point where the density of states becomes the highest and
the instantaneous energies change rapidly. In typical models,
adiabaticity is expected to be violated in the thermodynamic limit
$N\rightarrow\infty$, since then the curve crossing at the critical
point becomes infinitely sharp \cite{sachdev}. Here, we encounter a
different situations because already at the single particle level,
the energies at the crossing are degenerate making it a non-avoided
crossing. Thereby, the constrains for adiabaticity can be highly
tested already for small atom numbers, interpreted as small
scattering amplitudes $\mathbf{g}$.

In this section we analyze the excitations caused by non-adiabatic
effects. The time evolved wave packet reads
\begin{equation}\label{timesol}
\Psi(\mathbf{x},t)=\mathcal{T}e^{-i\int_0^t\hat{H}_{GP}(t')dt'}\Psi_{GS}(\mathbf{x}),
\end{equation}
where $\mathcal{T}$ is the time-ordering operator and
$\Psi_{GS}(\mathbf{x})$ is given in Eq.~(\ref{gs}). At any instant
of time we can apply the imaginary time evolution to find the
corresponding ground state $\Psi_{GS}^{(t)}(\mathbf{x})$ by simply
substitute $\Psi(\mathbf{x},t)$ into Eq.~(\ref{gs}). We introduce
the non-adiabatic relative energy increase as
\begin{equation}\label{aden}
\Delta E(t)=\frac{E_\Psi(t)-E_{\Psi_{GS}^{(t)}}(t)}{|E_{\Psi_{GS}^{(t)}}(t)|},
\end{equation}
where $E_\Psi(t)$ and $E_{\Psi_{GS}^{(t)}}(t)$ are the system
energies given by the functionals $E[\Psi]$ and $E[\Psi_{GS}^{(t)}]$
respectively. The condition $\partial\Delta E(t)/\partial t\ll1$ is
an indication that the evolution is adiabatic.

\begin{figure}[h]
\centerline{\includegraphics[width=7cm]{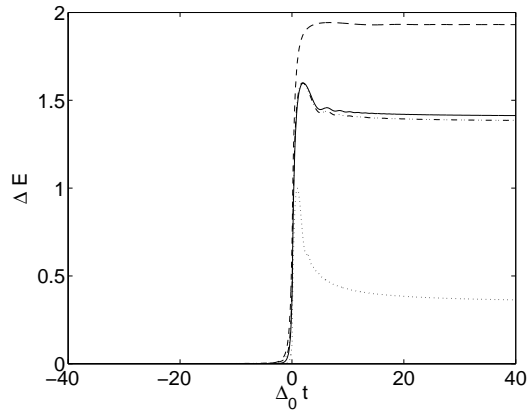}} \caption{The
relative energy increase (\ref{aden}) for different sweep velocities
(dotted line: $\Delta_0=1$, solid line and dot-dashed line:
$\Delta_0=10$, and dashed line: $\Delta_0=50$) and scattering
amplitudes (dotted/solid/dashed lines: ${\bf g}=0$ and dot-dashed
line: $g_{11}=g_{22}=10g_{12}=20$). Increasing $\Delta_0$ decreases
the adiabaticity as predicted. The solid and dash-dotted lines
correspond to the same sweep velocity but different scattering
parameters ${\bf g}$, and it is noted that the non-linearity does
not destroy adiabaticity in this parameter regime. However, for
larger non-linearity we have found great decrease of adiabaticity.
Note that we have scaled time with the sweep velocity on the
$x$-axis. The other dimensionless parameters are as in
Fig.~\ref{fig3} (a). } \label{fig8}
\end{figure}

We note that by neglecting the spatial variation of the optical
lattice, $\cos(x)\rightarrow1$, the Hamiltonian (\ref{ham3}) with
the detuning (\ref{timedet}) is identical to the Landau-Zener model
\cite{lz}. A fast sweep (large $\Delta_0$) implies non-adiabatic
evolution and a large change in $\Delta E(t)$. Another possible
reason for breakdown of adiabaticity is the non-linearity of the
Gross-Pitaevskii equation. It is known that strong non-linearity
(large values of ${\bf g}$) tends to decrease adiabaticity
\cite{niu}. On the other hand, for certain systems moderate
non-linearity can even increase adiabaticity \cite{nonads}. In
Fig.~\ref{fig8} we present examples of the relative energy, both for
different sweep velocities $\Delta_0$ and different strengths ${\bf
g}$. For the solid/dotted/dashed lines we use zero scattering ${\bf
g}=0$, but change the sweep velocity $\Delta_0$. The breakdown of
adiabaticity is dominantly occurring around the critical point as
expected, and moreover, fast sweeps gives a larger breakdown. The
case of a non-zero scattering is depicted as a dot-dashed line in
the figure. For this choice of scattering amplitudes
($g_{11}=g_{22}=10g_{12}=20$), the adiabaticity is approximately
unchanged. However, our numerical results indicates that around
$g_{11}=g_{22}=10g_{12}=30$ a drastic drop in the adiabaticity
appears. Such sudden decrease of adiabaticity has also been
demonstrated in related models~\cite{pethick}.

\section{Conclusion}\label{sec4}
We have presented a thorough numerical analysis of spinor BEC driven
through a critical point. The presence of a first order PT for the
interacting system was motivated by first studying the spectrum of
the single particle Hamiltonian. Since the spectrum is non-gapped,
the model could be sensitive to non-adiabatic excitations. Despite
this fact, it was shown by solving the full Gross-Pitaevskii
equation that the PT is still clearly visible even at relatively
fast sweeps through the critical point. The effects of atom-atom
collisions (Fig.~\ref{fig4}) as well as harmonic trapping
(Fig.~\ref{fig7}) were also analyzed, Both were believed to smear
out signatures of the PT. However, our findings demonstrate that the
PT is resilient to both. Due to limitation of computational power,
most results were carried out in the one dimensional situations, but
Fig.~\ref{fig6} made clear that the PT is not restricted to this
lower dimensional case only. As a concluding remark, the
demonstrated robustness of the PT makes the current model an
interesting system for experimental investigation.

\begin{acknowledgments}
JL acknowledges support from the MEC program (FIS2005-04627).
\end{acknowledgments}


\begin{thebibliography}{999}
\bibitem{bec} M. H. Anderson, J. R. Enscher, M. R. Matthews, C. E.
Wieman, and E. A. Cornell, Science {\bf 269}, 198 (1995); K. B.
Davies, M. O. Mewes, M. R. Andrews, N. J. Vandruten, D. S. Durfee,
D. M. Kurn, and W. Ketterle, Phys. Rev. Lett. {\bf 75}, 3969 (1995).

\bibitem{bec2} C. J. Pethick and H. Smith, {\it Bose-Einsten
Condensation in Dilute Gases}, (Cambridge University Press,
Cambridge 2008).

\bibitem{spinor} C. J. Myatt, E. A. Burt, R. W. Ghrist, E. A.
Cornell, and C. E. Wieman, Phys. Rev. Lett. {\bf 78}, 586 (1997); J.
Stenger, S. Inouye, D. M. Stamper-Kurn, H.-J. Miesner, A. P.
Chikkatur, and W. Ketterle, Nature {\bf 396}, 345 (1998); M. S. Chang, C. D. Hamley, M. D. Barrett, J. A. Sauer, K. M. Fortier, W. Zhang, L. You, and M. S. Chapman, Phys. Rev. Lett. {\bf 92}, 140403 (2004).

\bibitem{spintrans} O. Mandel, M. Greiner, A. Widera, T. Rom, T. W.
H\"ansch, and I. Bloch, Phys. Rev. Lett. {\bf 91}, 010407 (2003).

\bibitem{spinmix} M. S. Chang, C. D. Hamley, M. D. Barrett, J. A.
Sauer, K. M. Fortier, W. Zhang, L. You, and M. S. Chapman, Phys.
Rev. Lett. {\bf 92}, 140403 (2004).

\bibitem{spintun} D. M. Stamper-Kurn, H. J. Miesner, A. P.
Chikkatur, S. Inouye, J. Stenger, and W. Ketterle, Phys. Rev. Lett.
{\bf 83}, 661 (1999).

\bibitem{spinbreak} L. E. Sadler, J. M. Higbie, S. R. Leslie, M.
Vengalottore, and M. S. Stamper-Kurn, Nature {\bf 443}, 312 (2006).

\bibitem{bloch} O. Morsch, J. H. M\"uller, M. Cristiani, D.
Ciampini, and E. Arimondo, Phys. Rev. Lett. {\bf 87}, 140402 (2001).

\bibitem{gapsol} B. Eiermann, T. Anker, M. Albiez, M. Taglieber, P. Treutlein, K. P. Marzlin, M. K. Oberthaler, Phys. Rev. Lett. {\bf  92}, 230401 (2004).

\bibitem{solcol} Z. D. Li, P. B. He, J. Q. Liang, and W. M. Liu, Phys. Rev. A {\bf 71}, 053611 (2004).

\bibitem{selftrap} B. B. Wang, P. M. Fu, J. Liu, and B. Wu, Phys. Rev. A {\bf 74}, 063610 (2006).

\bibitem{super} J. Mun, P. Medley, G. K. Campbell, L. G. Marcassa,
D. E. Pritchard, and W. Ketterle, arXiv:0706.3946.

\bibitem{vortex} J. W. Reijnders and R. A. Duine, Phys. Rev. Lett.
{\bf 93}, 060401 (2004).

\bibitem{anderson} B. Juliette, V. Josse, Z. Zuo, A. Bernard, B.
Hembrecht, P. Lugan, D. Cl\'ement, L. Sanchez-Palencia, P. Bouyer,
and A. Aspect, Nature {\bf 453}, 891 (2008).

\bibitem{mott} M. Greiner, O. Mandel, T. Esslinger, T. W. H\"ansch,
and I. Bloch, Nature {\bf 415}, 39 (2002).

\bibitem{maciek} M. Lewenstein, A. Sanpera, V. Ahufinger, B. Damski, A. Sen, and U. Sen, Adv. Phys. {\bf 56}, 243 (2007); I. Bloch, J. Dalibard, and W. Zwerger, Rev. Mod. Phys. {\bf 80}, 885 (2008).

\bibitem{spintr0} W. H. Zurek, U. Dorner, and P. Zoller, Phys. Rev. Lett. {\bf 95}, 105701 (2005).

\bibitem{spintr1} J. Dziarmaga, Phys. Rev. Lett. {\bf 95}, 245701 (2005); T. Caneva, R. Fazio, and G. E. Santoro, Phys. Rev. B {\bf 76}, 144427 (2007); V. Mukherjee, U. Divakaran, A. Dutta, and D. Sen, Phys. Rev. B {\bf 76}, 174303 (2007); F. Pellegrini, S. Montangero, G. E. Santoro, and R. Fazio, Phys. Rev. B {\bf 77}, 140404(R) (2008).

\bibitem{spintr2} T. Caneva, R. Fazio, and G. E. Santoro, Phys. Rev. B {\bf 78}, 104426 (2008); P. Solinas, P. Riberio, and R. Mosseri, Phys. Rev. A {\bf 78}, 052329 (2008).

\bibitem{spintr3} K. Sengupta, D. Sen, and  S. Mondal, Phys. Rev. Lett. {\bf 100}, 077204 (2008); S. Mondal, D. Sen, and K. Sengupta, Phys. Rev. B {\bf 78}, 045101 (2008).

\bibitem{spintr4} H. T. Quan, Z. D. Wang, and C. P. Sun, Phys. Rev. A {\bf 76}, 012104 (2007)

\bibitem{kzbh1} A. Polkovnikov, Phys. Rev. B {\bf 72}, 161201(R) (2005); F. M. Cucchietti, B. Damski, J. Dziarmaga, and W. H. Zurek, Phys. Rev. A {\bf 75}, 023603 (2007).

\bibitem{kzbh2} J. Dziamaga, J. Meisner, and W. H. Zurek, Phys. Rev. Lett. {\bf 101}, 115701 (2008).

\bibitem{lz} L. D. Landau, Phys. Z. Sowjet Union {\bf 2}, (1932); C. Zener, Proc. Roy. Soc. Lond. A {\bf 137}, 696 (1932).

\bibitem{jonasjani} J. Larson and J.-P. Martikainen, Phys. Rev. A {\bf 78}, 063618 (2008).

\bibitem{jonasjani2} J. Larson and J.-P. Martikainen, arXiv:0811.4147.

\bibitem{adel} M. Alexanian and S. K. Bose, Phys. Rev. A {\bf 52}, 2218 (1995).

\bibitem{jonaseff} W. Ren and H. J. Carmichael, Phys. Rev. A {\bf 51}, 752 (1995);
J. Larson, J. Salo, and S. Stenholm, Phys. Rev. A {\bf 72}, 013814
(2005).

\bibitem{konstantin} K. V. Krutitsky and R. Graham, Phys. Rev. Lett. {\bf 91}, 240406 (2003); {\it ibid.}, Phys. Rev. A {\bf 70}, 063610 (2004).

\bibitem{pethick} B. Wu, R. B. Diener, and Q. Niu, Phys. Rev. A {\bf 65}, 025601 (2002); M. Machholm, C. J. Pethick, and H. Smith, Phys. Rev. A {\bf 67}, 053613 (2003).

\bibitem{sachdev} S. Sachdev, {\it Quantum Phase Transition},
(Cambdridge University Press, Cambridge 1998).

\bibitem{split} M. D. Fleit, J. A. Fleck, and A. Steiger, J. Comp. Phys. {\bf 47}, 412 (1982).

\bibitem{fidde0} W. L. You, Y. W. Li, and S.-J. Gu, Phys. Rev. E
{\bf 76}, 022101 (2007).

\bibitem{fidde1} S.-J. Gu, H.-M. Kwok, W.-Q. Ning, and H.-Q. Lin,
Phys. Rev. B {\bf 77}, 245109 (2008).

\bibitem{topfid} S. Yang, S.-J. Gu, C.-P. Sun, and H.-Q. Lin, Phys.
Rev. A {\bf 78}, 012304 (2008).

\bibitem{fidde2} H.-M. Kwok, C.-S. Ho, and S.-J. Gu, Phys. Rev. A {\bf 78}, 062302 (2008).

\bibitem{niu} B. Wu and Q. Niu, Phys. Rev. A {\bf 61}, 023402 (2000).

\bibitem{nonads} A. P. Itin and S. Watanabe, Phys. Rev. Lett. {\bf 99}, 223903 (2007).

\end{thebibliography}
\end{document}